\documentclass[structabstract]{aa}  
\usepackage{graphicx}
\usepackage{txfonts}
\usepackage{pstricks,pst-node,pst-plot}
\usepackage{natbib}
\bibpunct{(}{)}{;}{a}{}{,} 
\begin{document}
   \title{Comoving-frame radiative transfer in arbitrary velocity fields -- II.}
   \subtitle{Large scale applications.}


   \author{Sebastian Knop
          \inst{1}
          \and
          Peter H. Hauschildt
          \inst{1}
          \and
          Edward Baron
          \inst{1,2,3}
          }

   \institute{Hamburger Sternwarte, Gojenbergsweg 112, 21029 Hamburg, Germany\\
              \email{[sknop,yeti]@hs.uni-hamburg.de}
         \and
             Homer L.~Dodge Department of Physics and Astronomy, University of Oklahoma, 440 W Brooks, Rm 100, Norman, OK 73019 USA\\
             \email{baron@ou.edu}
         \and
        Computational Research Division, Lawrence Berkeley
        National Laboratory, MS 50F-1650, 1 Cyclotron Rd, Berkeley, CA
        94720 USA
             }

   \date{}

 
  \abstract
   {}
   {A solution of the radiative-transfer problem in arbitrary velocity fields
   introduced in a previous paper, has limitations in its applicability.
   For large-scale applications, the methods described also require large
   memory sets that are commonly not available to state-of-the-art computing
   hardware.  In this work, we modify the algorithm
   to allow the computation of large-scale problems.
   }
   {We reduce the memory footprint via a domain decomposition. By
   introducing iterative Gauss-Seidel type solvers, we improve the
   speed of the overall computation. Because of
   the domain decomposition, the new
   algorithm requires the use of parallel-computing systems.
   }
   {
   The algorithm that we present permits large-scale solutions of
   radiative-transfer problems that include arbitrary wavelength couplings.  In
   addition, we discover a quasi-analytic formal solution of the radiative
   transfer that significantly improves the overall computation speed. More
   importantly, this method ensures that our algorithm can be applied to
   multi-dimensional Lagrangian radiative-transfer calculations. In
   multi-dimensional atmospheres, velocity fields are in general chaotic
   ensuring that the inclusion of arbitrary wavelength couplings are mandatory.
   }
   {}


   \maketitle
%
\section{Introduction}
\label{sec:01}

Radiative transfer in spherical symmetry and moving media has been solved
with different methods. Operator-splitting techniques \citep{cannon1973ApJ}
remain the state-of-the-art methods. An operator-splitting method that uses
an approximate $\Lambda$-operator (ALO) \citep{scharmer1981} -- called 
accelerated $\Lambda$-iteration (ALI) -- for the solution of the
special relativistic Lagrangian equation of radiative transfer was
described, for example, in \cite{peter1992JQSRT}.  A Lagrangian description is
necessary for the treatment of line transfer because of the
complicated form of the emissivities and opacities in the Eulerian frame and also
the large number of wavelength points needed to sample the
spectral line at all points in the atmosphere. It follows that the velocity field
must then appear explicitly in the equation of radiative transfer \citep{mihalas80}.
For monotonic velocity fields, the solution of the radiative-transfer problem
becomes an initial value problem in wavelength that can be solved with the method described in
\cite{peter1992JQSRT}. The inclusion of non-monotonic velocity fields changes
the problem into a boundary-value problem in wavelength. A solution method and its
algorithm for this case was introduced in \cite{petereddienms3}
(Paper~I).

However, the applicability of the methods in Paper~I depends on the size of the
numerical system, which is determined mainly by the number of wavelength points as
well as the number of layers in the 1-D model atmosphere.  The number of
wavelength points is the more critical factor, since it is commonly much higher than
the number of layers in 1-D.  The number of layers is however important, because
the number of entries in the $\Lambda$-operator described in Paper~I 
is proportional to the number of layers squared.

On average hardware, the 
algorithm from Paper~I is only suited to solving problems with 64 layers
and about 1000 wavelength points because of the limited
memory available. Configurations with more than 1000 wavelength points are
called in the following, large-scale applications.

The work from Paper~I must be regarded as a proof of concept. In this work,
we present a new algorithm that is capable of efficiently solving large-scale
applications. Besides being immediately useful for 1-D applications,
the results are also applicable to the development of 3-D radiative transfer \citep{peter3d_I,peter3d_II,peter3d_III}.

In Sect.~\ref{sec:02}, we summarise the formalism again for completeness.
The approach for the new algorithm is
described in Sect.~\ref{sec:03}. In Sect.~\ref{sec:04}, we present the results for
test calculations and scalability tests before we conclude in Sect.~\ref{sec:05}

\section{The formal framework in general}
\label{sec:02}

The framework presented in \cite{petereddienms3} was developed for just
one possible wavelength discretisation.  During the course of this work, it
became clear that a fully implicit discretisation \citep{petereddie2004} is
necessary to solve the radiative-transfer problem in all generality.  This is
because for strong wavelength couplings the interpretation of the
entire wavelength discretisation as a source term is not valid for all optical
depths.
Furthermore, the use of a new formal solution that avoids negative generalised
opacities must be used \citep{newfrmsol}. Therefore, we discuss here the basic
framework in those parts that are different from \cite{petereddienms3}.

The equation of radiative transfer in its characteristic form for the specific
intensity $I$ along a path $s$ reads:
\begin{equation}
\label{eq:eqrt}             
\frac{\mathrm{d} I_l}{\mathrm{d} s} = \eta_l - \chi_l I_l - 4 a_l I_l - a_l \frac{\partial(\lambda I_l)}{\partial \lambda} ,
\end{equation}
where $\eta$ is the emissivity, $\chi$ the opacity, and the subscript $l$
indicates the dependence of the intensity on wavelength. The $a_l$ term is the coupling term between
the wavelengths that depends on both the structure of the atmosphere and the
mechanism of the coupling \citep{mihalas80}.

The wavelength derivative can be discretised in two ways as described in
\cite{petereddie2004}. The different discretisations can be mixed via a
Crank-Nicholson scheme with a mixing parameter $\xi \in [0,1]$. The wavelength-discretised 
equation of radiative transfer can then be written as:
\begin{eqnarray}
\label{eq:eqrt2}
\frac{\mathrm{d} I_l}{\mathrm{d} s} &=& \eta_l - \chi_l I_l - a_l \left(4 + \xi \;  p_l^| \right) I_l \nonumber \\ 
                                    &{}& \;  - \xi \; a_l \left( p_l^- I_{{l-1}} + p_l^+ I_{{l+1}}  \right)    \nonumber \\
				    &{}& \;  - [1 - \xi] \;  a_l \left(   p_l^- I_{{l-1}} + p_l^| I_l + p_l^+ I_{{l+1}}      \right)
\end{eqnarray}
where the $p_l^\bullet$ coefficients in an ordered wavelength grid $\lambda_{l-1}
< \lambda_l < \lambda_{l+1}$ are defined as:
\begin{eqnarray}
\left.
\begin{array}{r c l}
 \displaystyle p_l^- & = & \displaystyle - \frac{\lambda_{l-1}}{\lambda_l - \lambda_{l-1}} \\
 \displaystyle p_l^| & = & \displaystyle \phantom{-}  \frac{\lambda_{l}}{\lambda_l - \lambda_{l-1}}  \\
 \displaystyle p_l^+ & = & \displaystyle \phantom{-} 0
\end{array}
\right\} & \mathrm{for:~} & a_{\lambda_l} \ge 0 \\
\left.
\begin{array}{r c l}
 \displaystyle p_l^- & = & \displaystyle \phantom{-} 0 \\
 \displaystyle p_l^| & = & \displaystyle \phantom{-}  \frac{\lambda_{l}}{\lambda_l - \lambda_{l+1}} \\ 
 \displaystyle p_l^+ & = & \displaystyle - \frac{\lambda_{l+1}}{\lambda_l - \lambda_{l+1}}
\end{array}
\right\} & \mathrm{for:~} & a_{\lambda_l} < 0 
\end{eqnarray}
The dependence on the sign of $a_\lambda$ is introduced to define local upwind
schemes (see \cite{petereddienms3}).

After introducing a generalised opacity (see \cite{newfrmsol})
\begin{equation}
\label{eq:chihat}
\hat{\chi}_l = \chi_l + \xi \; a_l p_l^| ,
\end{equation}
as well as defining the source functions:
\begin{eqnarray}
S_l & = & \frac{\eta_l}{\chi_l} \\
\hat{S}_l & = & \frac{\chi_l}{\hat{\chi}_l} \left(  S_l - \xi \; \frac{a_l}{\chi_l} \left( p_l^- I_{{l-1}} + p_l^+ I_{{l+1}} \right)  \right) \label{eq:srchat} \\
\tilde{S}_l & = & -\, \frac{a_l}{\hat{\chi}_l}  \left( [1 - \xi] \;  \left(  p_l^- I_{{l-1}} +  p_l^+ I_{{l+1}}\right) + \left[ 4 + [1 - \xi] \; p_l^| \right]  I_l \right) \label{eq:srctilde}
\end{eqnarray}
a formal solution of the radiative-transfer problem can be formulated. In our
work, we use a characteristic method along different photon paths throughout the
atmosphere. The spatial position of a characteristic is then discretised on a
spatial grid. In the following, a pair of subscript indices indicate the
position in both the spatial grid and the wavelength grid.  Commonly the spatial
grid is mapped onto an optical depth grid via the relation $\mathrm{d} \tau_l =
\hat{\chi}_l \mathrm{d} s$.  The formal solution of the equation of radiative
transfer in Eq.~\ref{eq:eqrt2} between two points $s_{i-1}$ and $s_{i}$ on a spatial
grid along the photon path can be written in terms of the optical depth  as follows:
\begin{eqnarray}
\label{eq:frmsl}
I_{i,l} & = & I_{i-1,l} e^{- \Delta \tau} +  \delta \hat{I}_{i,l} +  \delta \tilde{I}_{i,l} \\
\delta \hat{I}_{i,l}   & = &\int_{\tau_{i-1}}^{\tau_{i}} \! \! \!  \hat{S}_l  e^{\tau - \tau_{i}} \mathrm{d} \tau = \alpha_{i,l} \hat{S}_{i-1,l} + \beta_{i,l} \hat{S}_{i,l} + \gamma_{i,l} \hat{S}_{i+1,l} \label{eq:deltahat}\\
\delta \tilde{I}_{i,l} & = &\int_{\tau_{i-1}}^{\tau_{i}} \! \! \!  \tilde{S}_l  e^{\tau - \tau_{i}} \mathrm{d} \tau = \tilde{\alpha}_{i,l} \tilde{S}_{i-1,l} + \tilde{\beta}_{i,l} \tilde{S}_{i,l} \label{eq:deltatilde}
\end{eqnarray}
with $\Delta \tau = \tau_{i+1,l} - \tau_{i,l}$ and $\tau_{i,l} = \int^{s_i}
\chi_l(s) \mathrm{d} s$. The $\alpha$-$\beta$-$\gamma$ coefficients are
described in \cite{frmsol2} and \cite{peter1992JQSRT}. In
Eq.~\ref{eq:deltatilde}  $\delta \tilde{I}_l$ is interpolated linearly and the coefficients differ in general from
the coefficients in Eq.~\ref{eq:deltahat}, and is therefore marked with a tilde.

Equation~\ref{eq:frmsl} can be written in matrix notation for any given
characteristic:
\begin{equation}
\label{eq:frml_sln_matrix}
\mathbf{I} = \mathbf{A} \cdot \mathbf{I} + \mathbf{\Delta {I}}
\end{equation}
Here $\mathbf{I}$ is a vector containing all intensities, $\mathbf{A}$ is a square matrix
that describes the influence of the different intensities upon each other, and
$\mathbf{\Delta {I}}$ is a vector with the thermal emission and scattering
contribution of the source function.  For a characteristic with $n_i$ spatial
points and $n_l$ points in the wavelength grid, the intensity vector
$\mathbf{I}$ has $n_i \times n_l$ entries.  In the following, a superscript of
$k$ labels the characteristic being described. The components of the matrix
$\mathbf{A}$ from Eq.~\ref{eq:frml_sln_matrix} at the spatial point $i$ and the
wavelength point $l$ are given by:

\begin{eqnarray}
A^{\mathrm{-},k}_{i,l} & = & - \left( \xi \alpha^k_{i,l} +  [1 - \xi] \, \tilde{\alpha}^k_{i,l} \right) \frac{a^k_{i-1,l}}{\hat{\chi}^k_{i-1,l}}  p^{-,k}_{i-1,l} \label{eq:matrix_coeff_start} \\
B^{\mathrm{-},k}_{i,l} & = & - \left( \xi \beta^k_{i,l} +  [1 - \xi] \, \tilde{\beta}^k_{i,l} \right) \frac{a^k_{i,l  }}{\hat{\chi}^k_{i,l  }}    p^{-,k}_{i,l} \\
C^{\mathrm{-},k}_{i,l} & = &- \xi \gamma^k_{i,l} \frac{a^k_{i+1,l}}{\hat{\chi}^k_{i+1,l}} p^{-,k}_{i+1,l}\\
A^{\mathrm{\diagdown},k}_{i,l} & = & \! \! \exp{(-\Delta \tau^k_{i-1,l})} - \tilde{\alpha}^k_{i,l} \frac{a^k_{i-1,l}}{\hat{\chi}^k_{i-1,l}} \left[ 4 + [1\!-\! \xi]  p^{|,k}_{i-1,l} \right] \\
B^{\mathrm{\diagdown},k}_{i,l} & = &- \tilde{\beta}^k_{i,l} \frac{a^k_{i,l}}{\hat{\chi}^k_{i,l}} \left[ 4 + [1 - \xi] \, p^{|,k}_{i,l} \right]\\
C^{\mathrm{\diagdown},k}_{i,l} & = &- \hat{\gamma}^k_{i,l} \frac{a^k_{i+1,l}}{\hat{\chi}^k_{i+1,l}} p^{|,k}_{i+1,l}\\
C^{\mathrm{\diagdown},k}_{i,l} & = & 0 \\
A^{\mathrm{+},k}_{i,l} & = &- \left( \xi \alpha^k_{i,l} +  [1 - \xi] \, \tilde{\alpha}^k_{i,l} \right) \frac{a^k_{i-1,l}}{\hat{\chi}^k_{i-1,l}}  p^{+,k}_{i-1,l} \label{eq:matrix_super1} \\
B^{\mathrm{+},k}_{i,l} & = &- \left( \xi \beta^k_{i,l} +  [1 - \xi] \, \tilde{\beta}^k_{i,l} \right) \frac{a^k_{i,l  }}{\hat{\chi}^k_{i,l  }}    p^{+,k}_{i,l} \label{eq:matrix_super2} \\
C^{\mathrm{+},k}_{i,l} & = &- \xi \gamma^k_{i,l} \frac{a^k_{i+1,l}}{\hat{\chi}^k_{i+1,l}} p^{+,k}_{i+1,l} \label{eq:matrix_coeff_end}
\end{eqnarray}
The naming scheme of the quantities defined in Eqs.~\ref{eq:matrix_coeff_start} to
\ref{eq:matrix_coeff_end} indicates the specific intensity element with which they
are associated. For an index pair $i$ and $l$, a $\bullet^{\mathrm{-}}$
superscript refers to an intensity at wavelength $l-1$, a
$\bullet^{\mathrm{\diagdown}}$ superscript to the same wavelength, and
$\bullet^{\mathrm{+}}$ to the next wavelength point $l+1$. The $A, B$, and $C$ terms
refer to the spatial points $i-1, i$, and $i+1$ respectively.  For clarity, the
structure of the matrix of the formal solution is shown schematically in
Fig.~\ref{fig:frml_sln_matrix}.

\begin{figure*}
\begin{center}
\begin{minipage}{0.85\hsize}
\begin{displaymath}
\arraycolsep=1pt
\left( \begin{array}{c}
I_{1,1} \\
\vdots \\
\vdots \\
\vdots \\
I_{1,k}\\ \hline
I_{2,1}\\
\vdots \\
\vdots \\
\vdots \\
I_{2,k} \\ \hline
\vdots \\
\vdots \\
\vdots \\ \hline
I_{l,1} \\
\vdots \\
\vdots \\
\vdots \\
I_{l,k}
\end{array} \right)
=
\arraycolsep=2pt
\left( \begin{array}{c c c c c c c c c c c c c c c c c c c }
B^{\diagdown} &  0  & 0 &\multicolumn{2}{c}{\stackrel{(k-2)}{\ldots \ldots} } & 0 & B^{+} & C^{+} & 0 & \multicolumn{9}{c}{\stackrel{((l-1)\times k) -2}{\ldots \ldots \ldots \ldots \ldots \ldots \ldots \ldots \ldots \ldots \ldots \ldots \ldots \ldots}} &0 \\
A^{\diagdown} & B^{\diagdown} &  0  & 0 & \stackrel{(k-3)}{\ldots} & 0 & A^{+} & B^{+} & C^{+} & 0 &  \multicolumn{8}{c}{\stackrel{((l-1)\times k) -3}{\ldots \ldots \ldots \ldots \ldots \ldots \ldots \ldots \ldots \ldots \ldots \ldots \ldots}}  &0 \\ 
0 & A^{\diagdown} & B^{\diagdown} &  0  & 0 & \stackrel{(k-3)}{\ldots} & 0 & A^{+} & B^{+} & C^{+} & 0 & \multicolumn{7}{c}{\stackrel{((l-1)\times k) -4}{\ldots \ldots \ldots \ldots \ldots \ldots \ldots \ldots \ldots \ldots \ldots \ldots}}  & 0 \\
\vdots & \ddots       &  \ddots           &       \ddots      & \ddots & \ddots & \ddots & \ddots & \ddots & \ddots & \ddots & \ddots & \ddots & \ddots & \ddots & \ddots & \ddots & \ddots & \vdots \\
0 & \stackrel{(k-2)}{\ldots} & 0 & A^{\diagdown} & B^{\diagdown} & 0 & \stackrel{(k-2)}{\ldots} & 0 & A^{+} & B^{+} & 0 & \multicolumn{7}{c}{\stackrel{(l-2)\times k}{\ldots \ldots \ldots \ldots \ldots \ldots \ldots \ldots \ldots \ldots \ldots \ldots}}    &0\\ \hline
B^{-} & C^{-} & 0 &\multicolumn{2}{c}{\stackrel{(k-2)}{\ldots \ldots}}  & 0 & B^{\diagdown} &  0  & 0 & \multicolumn{2}{c}{\stackrel{(k-2)}{\ldots \ldots}} & 0 & B^{+} & C^{+} & 0 & \multicolumn{3}{c}{\stackrel{((l-2)\times k)-2}{\ldots \ldots  \ldots \ldots \ldots }}   &0 \\
A^{-} & B^{-} & C^{-} & 0 &\stackrel{(k-3)}{\ldots}  & 0 &  A^{\diagdown} & B^{\diagdown} &  0  & 0 & \stackrel{(k-3)}{\ldots} & 0 & A^{+} &  B^{+} & C^{+} & 0 & \multicolumn{2}{c}{\stackrel{((l-2) \times k)-3}{\ldots \ldots}} & 0 \\
0 & A^{-} & B^{-} & C^{-} & 0 &\stackrel{(k-3)}{\ldots}  & 0 &  A^{\diagdown} & B^{\diagdown} &  0  & 0 & \stackrel{(k-3)}{\ldots} & 0 & A^{+} &  B^{+} & C^{+} & 0&  \stackrel{((l-2) \times k)-4 }{\ldots \ldots} & 0 \\
\vdots & \ddots       &  \ddots           &       \ddots      & \ddots & \ddots & \ddots & \ddots & \ddots & \ddots & \ddots & \ddots & \ddots & \ddots & \ddots & \ddots & \ddots & \ddots & \vdots \\
0 & \stackrel{(k-2)}{\ldots} & 0& A^{-} & B^{-} & C^{-} & 0 & \stackrel{(k-2)}{\ldots} & 0 &  A^{\diagdown} & B^{\diagdown} & 0 & \stackrel{(k-2)}{\ldots} & 0 & A^{+} & B^{+} & 0 & \stackrel{{\scriptscriptstyle{((l-3) \times k)} }}{\ldots} & 0\\ \hline
\vdots & \ddots       &  \ddots           &       \ddots      & \ddots & \ddots & \ddots & \ddots & \ddots & \ddots & \ddots & \ddots & \ddots & \ddots & \ddots & \ddots & \ddots & \ddots & \vdots \\
\vdots & \vdots       &  \vdots           &       \vdots      & \vdots & \vdots & \vdots & \vdots & \vdots & \vdots & \vdots & \vdots & \vdots &\vdots & \vdots & \vdots & \vdots & \vdots & \vdots \\
\vdots & \ddots       &  \ddots           &       \ddots      & \ddots & \ddots & \ddots & \ddots & \ddots & \ddots & \ddots & \ddots & \ddots & \ddots & \ddots & \ddots & \ddots & \ddots & \vdots \\ \hline
0& \multicolumn{5}{c}{ \stackrel{(l-2)\times k}{\ldots \ldots \ldots \ldots \ldots \ldots \ldots \ldots}   }&0&B^{-} & C^{-} &0&\multicolumn{2}{c}{ \stackrel{(k-2)}{\ldots\ldots}  } &0&B^{\diagdown} &  0  & 0 & \multicolumn{2}{c}{\stackrel{(k-2)}{\ldots\ldots}} & 0 \\
0& \multicolumn{5}{c}{\stackrel{(l-2)\times k}{\ldots \ldots \ldots \ldots \ldots \ldots \ldots \ldots}}&0&A^{-} &B^{-} & C^{-} &0&  \stackrel{(k-3)}{\ldots}  &0&A^{\diagdown} &B^{\diagdown} &  0  & 0 & \stackrel{(k-3)}{\ldots} & 0 \\
\vdots & \ddots       &  \ddots           &       \ddots      & \ddots & \ddots & \ddots & \ddots & \ddots & \ddots & \ddots & \ddots & \ddots & \ddots & \ddots & \ddots & \ddots & \ddots & \vdots \\
0& \multicolumn{8}{c}{\stackrel{((l-1)\times k) -2}{\ldots \ldots \ldots \ldots \ldots \ldots \ldots \ldots}}&0&A^{-} & B^{-} &C^{-} &0&\stackrel{(k-3)}{\ldots} &0&A^{\diagdown} & B^{\diagdown}&  0  \\
0& \multicolumn{9}{c}{\stackrel{((l-1)\times k) -2}{\ldots \ldots \ldots \ldots \ldots \ldots \ldots \ldots} }&0&A^{-} & B^{-} &0& \multicolumn{2}{c}{ \stackrel{(k-2)}{\ldots \ldots}}  &0&A^{\diagdown} & B^{\diagdown}
\end{array}
\right) \cdot 
\left( \begin{array}{c}
I_{1,1} \\
\vdots \\
\vdots \\
\vdots \\
I_{1,k}\\ \hline
I_{2,1}\\
\vdots \\
\vdots \\
\vdots \\
I_{2,k} \\ \hline
\vdots \\
\vdots \\
\vdots \\ \hline
I_{l,1} \\
\vdots \\
\vdots \\
\vdots \\
I_{l,k}
\end{array} \right) +
\left( \begin{array}{c}
\arraycolsep0pt
\Delta \hat{I}_{1,1} \\
\vdots \\
\vdots \\
\vdots \\
\Delta \hat{I}_{1,k}\\ \hline
\Delta \hat{I}_{2,1}\\
\vdots \\
\vdots \\
\vdots \\
\Delta \hat{I}_{2,k} \\ \hline
\vdots \\
\vdots \\
\vdots \\ \hline
\Delta \hat{I}_{l,1} \\
\vdots \\
\vdots \\
\vdots \\
\Delta \hat{I}_{l,k}
\end{array} \right)   
\end{displaymath}
\end{minipage}
\caption[{Explicit form of the formal solution matrix}]{Explicit matrix form of
the formal solution for a characteristic of length $k$ and $l$ wavelength
points. The horizontal lines mark block borders of different wavelengths to
clarify the structure. 

The matrix has three tridiagonal bands. The one on the main diagonal is called
\emph{diag}$(=\diagdown)$ and the lower and upper accordingly \emph{sub}$(=-)$
and \emph{super}$(=+)$.  The diagonals of these bands are called \emph{A},
\emph{B}, and \emph{C}.}
\label{fig:frml_sln_matrix}
\end{center}
\end{figure*}

An element of the source function vector $\mathbf{\Delta {I}}$ is given by:
\begin{equation}
\Delta {I}^k_{i,l}   =  \alpha^k_{i,l} {S}^k_{i-1,l} + \beta^k_{i,l} {S}^k_{i,l} + \gamma^k_{i,l} {S}^k_{i+1,l} \\
\end{equation}
From Eq.~\ref{eq:frml_sln_matrix}, the solution for the specific intensity at a
given spatial point and wavelength reads:
\begin{eqnarray}
I^k_{i,l} & =  & \left(1 - B^{\mathrm{diag},k}_{i,l} \right)^{-1} \cdot \big( \Delta {I}^{k}_{i,l} +  B^{\mathrm{sub},k}_{i,l} I^k_{i,l-1} + B^{\mathrm{super},k}_{i,l} I^k_{i,l+1} \nonumber \\
 & & \quad  + A^{\mathrm{sub},k}_{i,l} I^k_{i-1,l-1} + A^{\mathrm{diag},k}_{i,l} I^k_{i-1,l}+  A^{\mathrm{super},k}_{i,l} I^k_{i-1,l+1} \nonumber \\
 & & \quad  + C^{\mathrm{sub},k}_{i,l} I^k_{i+1,l-1} + \phantom{C^{\mathrm{diag},k}_{i,l} I^k_{i+1,l}+}  C^{\mathrm{super},k}_{i,l} I^k_{i+1,l+1} \big) \label{eq:explicit_frml_sln0} 
\end{eqnarray}

Given the form of Eq.~\ref{eq:explicit_frml_sln0} for the formal solution, the
construction of the $\Lambda^\ast$-operator can proceed exactly as described in
\cite{petereddienms3} and will not be discussed further here.

\section{Optimisation of the algorithm}
\label{sec:03}

In the following sections, we outline the changes made to the algorithm from Paper~I to
improve the performance and usability of the solution to the radiative-transfer
problem in the case of large-scale applications. 

\subsection{Smaller amount of memory requirements}
\label{sec:03:00}

The solution of the equation of transfer for arbitrary wavelength couplings is a
boundary-value problem in wavelength \citep{petereddienms3},
where the wavelength derivative sense changes throughout the atmosphere. 
This implies that the radiative transfer must be
solved for all discrete wavelength points at the same time, and 
that all wavelength dependent quantities such as the opacities, interpolation
coefficients, wavelength-derivative discretisation, and the $\Lambda$-operator
must be kept in memory at the same time. For large-scale applications, these requirements easily exceed the
memory of commonly available computer hardware.  Therefore, the key method of solution is a
domain decomposition of the data. Ideally, every process stores only 
the data that it
works on. This immediately implies parallelisation of code execution as well.

The formal solution can be parallelised. The formal solutions for
different characteristics are independent of each other and accordingly a
computing node in a parallel setup must store only the data for
those characteristics upon which it works.

The $\Lambda^\ast$-operator is the largest data object that must be retained in
memory, and also offers the most hopeful possibility for optimisation. The operator has
the full spatial bandwidth but is only tridiagonal in wavelength. Therefore, the
number of entries of the $\Lambda^\ast$-operator is $n_{\mathrm{layer}} \times
n_{\mathrm{layer}} \times n_{\lambda} \times 3$, where $n_{\mathrm{layer}}$ is
the number of discrete radial points in a spherically symmetric atmosphere and
$n_{\lambda}$ is the number of discrete wavelength points.

For a model atmosphere with 100 layers and 20~000 wavelength points, the
operator takes up $\approx 4.5$~GB of memory. This easily exceeds the average
memory per processor available in modern computing systems. The complete
operator must be kept in memory for the solution of the ALI step if direct
solvers such as the LAPACK package are used.
The need to store the factorisation actually doubles the memory requirements.
However, this is not the case
if an iterative method of solution is used for the ALI step. Then different
tasks can work on different parts of the index field that are iterated. This
in turn means that only those parts of the operator that are needed for the local
iteration of the new mean intensities need to exist in each task. The
storage requirements are then reduced by a factor equal to the number of
tasks involved in the computation.

A drawback of this strategy of decomposing the operator
is that it greatly
increases the need for communication between tasks. The formal solution and
its accompanying data are parallelised over the characteristics of the radiation
field (see Sect.~\ref{sec:03:02}). A $\Lambda^\ast$-operator element is influenced by all
characteristics and therefore contributions to the element that is stored for just one
process must be calculated by all processes and be communicated. A
parallelised iterative solution of the mean intensity in the ALO step is
also enforced, increasing the need for communication even more (see Sect.~\ref{sec:03:03}).

\subsection{Optimisation of the speed of the formal solution}
\label{sec:03:01}

To improve the overall computational speed, the
time take to complete a formal solution must be decreased because it is
performed most often in an ALI solution.
In small-scale applications, the SuperLU package
\citep{superlu99} provided an efficient solver for the matrix equation of the
formal solution along a characteristic. 
However, there was room for improvement, e.g., in terms of memory footprint, so
we developed an iterative Gauss-Seidel (GS) type solver \citep{golub89}, which proved to
have a minimal memory footprint as well as to be very fast. The main advantage
of this new solver is, however,
that for a linearly interpolated source function $\hat{S}$, the formal solution,
becomes quasi-analytic and there is no need at all to solve a matrix equation.

In principle, our method is a standard GS type iterative solver that uses a physically
guided index field stepping scheme. That means that we use knowledge
of the physics along the characteristic to increase the convergence
of the iteration by a huge factor. The GS method does not prescribe the order in which the
elements of the system are iterated. Therefore, we can freely
choose the order of the steps
in the solution of the linear system. We choose it
so that we
follow the motion of a pulse along the characteristic, hence follow the physical flow
of information in the system. Because we know that the information will always
be propagated along the characteristic, we are left only with the task of determining whether the
information flows to longer or shorter wavelengths at any given spatial point.
This problem has been already solved in the construction of the approximate
$\Lambda$-operator and its solution can be reused here.

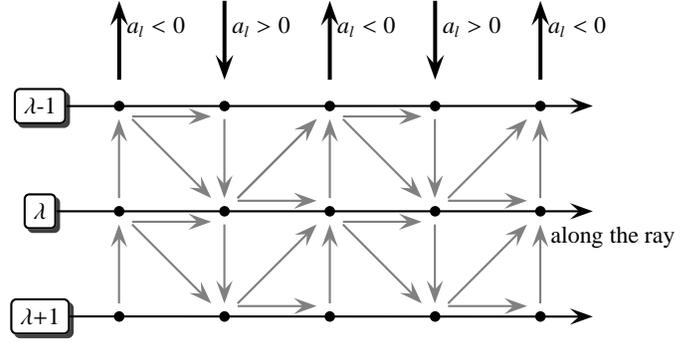
\begin{figure}
\begin{center}
	\psset{unit=0.7cm}
	\begin{pspicture}(-6.0,-2.5)(6.0,4.25)
        \rput(-5.5,2){\rnode{l-1}{\psframebox[framearc=0.3,cornersize=relative,shadow=true,shadowsize=0.1]{$\lambda$-1}}}
        \rput(-5.5,0){\rnode{l}{\psframebox[framearc=0.3,cornersize=relative,shadow=true,shadowsize=0.1]{$\lambda$}}}
        \rput(-5.5,-2){\rnode{l+1}{\psframebox[framearc=0.3,cornersize=relative,shadow=true,shadowsize=0.1]{$\lambda$+1}}}
        \rput(5,2){\rnode{l-1end}{}}
        \rput(5,0){\rnode{lend}{}}
        \rput(5,-2){\rnode{l+1end}{}}
        \ncline{->,arrowscale=2}{l-1}{l-1end}       
        \ncline{->,arrowscale=2}{l}{lend}       
        \ncline{->,arrowscale=2}{l+1}{l+1end}      
        \pscircle*(0,0){0.1}
        \pscircle*(0,2){0.1}
        \pscircle*(0,-2){0.1}
        \pscircle*(-2.0,0){0.1}
        \pscircle*(-2.0,2){0.1}
        \pscircle*(-2.0,-2){0.1}
        \pscircle*(-4.0,0){0.1}
        \pscircle*(-4.0,2){0.1}
        \pscircle*(-4.0,-2){0.1}
        \pscircle*(2,0){0.1}
        \pscircle*(2,2){0.1}
        \pscircle*(2,-2){0.1}
        \pscircle*(4,0){0.1}
        \pscircle*(4,2){0.1}
        \pscircle*(4,-2){0.1}
%
        \rput(-4,-2.0){  \rnode{-4-2}{}  }
        \rput(-2,-2.0){  \rnode{-2-2}{}  }
        \rput(0,-2.0){  \rnode{0-2}{}  }
        \rput(2,-2.0){  \rnode{2-2}{}  }
        \rput(4,-2.0){  \rnode{4-2}{}  }
        
        \rput(-4,-1.8){  \rnode{-4-2up}{}  }
        \rput(-2,-1.8){  \rnode{-2-2up}{}  }
        \rput(0,-1.8){  \rnode{0-2up}{}  }
        \rput(2,-1.8){  \rnode{2-2up}{}  }
        \rput(4,-1.8){  \rnode{4-2up}{}  }
        
	\rput(-4,0){  \rnode{-40}{}  }
        \rput(-2,0){  \rnode{-20}{}  }
        \rput(0,0){  \rnode{00}{}  }
        \rput(2,0){  \rnode{20}{}  }
        \rput(4,0){  \rnode{40}{}  }

        \rput(-4,0.2){  \rnode{-40up}{}  }
        \rput(-2,0.2){  \rnode{-20up}{}  }
        \rput(0,0.2){  \rnode{00up}{}  }
        \rput(2,0.2){  \rnode{20up}{}  }
        \rput(4,0.2){  \rnode{40up}{}  }
        \rput(-4,-0.2){  \rnode{-40dwn}{}  }
        \rput(-2,-0.2){  \rnode{-20dwn}{}  }
        \rput(0,-0.2){  \rnode{00dwn}{}  }
        \rput(2,-0.2){  \rnode{20dwn}{}  }
        \rput(4,-0.2){  \rnode{40dwn}{}  }
        
	\rput(-4,2){  \rnode{-42}{}  }
        \rput(-2,2){  \rnode{-22}{}  }
        \rput(0,2){  \rnode{02}{}  }
        \rput(2,2){  \rnode{22}{}  }
        \rput(4,2){  \rnode{42}{}  }

	\rput(-4,1.8){  \rnode{-42dwn}{}  }
        \rput(-2,1.8){  \rnode{-22dwn}{}  }
        \rput(0,1.8){  \rnode{02dwn}{}  }
        \rput(2,1.8){  \rnode{22dwn}{}  }
        \rput(4,1.8){  \rnode{43dwn}{}  }
%
	\ncline{->,arrowscale=2.0,nodesepA=0.25,nodesepB=0.25,linecolor=gray}{-4-2}{-40}
	\ncline{->,arrowscale=2.0,nodesepA=0.25,nodesepB=0.25,linecolor=gray}{-40}{-42}
	\ncline{->,arrowscale=2.0,nodesepA=0.25,nodesepB=0.25,linecolor=gray}{-40dwn}{-20dwn}
	\ncline{->,arrowscale=2.0,nodesepA=0.25,nodesepB=0.25,linecolor=gray}{-42dwn}{-22dwn}
	\ncline{->,arrowscale=2.0,nodesepA=0.25,nodesepB=0.25,linecolor=gray}{-22}{-20}
	\ncline{->,arrowscale=2.0,nodesepA=0.25,nodesepB=0.25,linecolor=gray}{-42}{-20}
	\ncline{->,arrowscale=2.0,nodesepA=0.25,nodesepB=0.25,linecolor=gray}{-20}{-2-2}
	\ncline{->,arrowscale=2.0,nodesepA=0.25,nodesepB=0.25,linecolor=gray}{-40}{-2-2}
	\ncline{->,arrowscale=2.0,nodesepA=0.25,nodesepB=0.25,linecolor=gray}{-20up}{00up}
	\ncline{->,arrowscale=2.0,nodesepA=0.25,nodesepB=0.25,linecolor=gray}{-2-2up}{0-2up}
	\ncline{->,arrowscale=2.0,nodesepA=0.25,nodesepB=0.25,linecolor=gray}{-2-2}{00}
	\ncline{->,arrowscale=2.0,nodesepA=0.25,nodesepB=0.25,linecolor=gray}{0-2}{00}
	\ncline{->,arrowscale=2.0,nodesepA=0.25,nodesepB=0.25,linecolor=gray}{00}{02}
	\ncline{->,arrowscale=2.0,nodesepA=0.25,nodesepB=0.25,linecolor=gray}{-20}{02}
	\ncline{->,arrowscale=2.0,nodesepA=0.25,nodesepB=0.25,linecolor=gray}{00dwn}{20dwn}
	\ncline{->,arrowscale=2.0,nodesepA=0.25,nodesepB=0.25,linecolor=gray}{02dwn}{22dwn}
	\ncline{->,arrowscale=2.0,nodesepA=0.25,nodesepB=0.25,linecolor=gray}{02}{20}
	\ncline{->,arrowscale=2.0,nodesepA=0.25,nodesepB=0.25,linecolor=gray}{22}{20}
	\ncline{->,arrowscale=2.0,nodesepA=0.25,nodesepB=0.25,linecolor=gray}{20}{2-2}
	\ncline{->,arrowscale=2.0,nodesepA=0.25,nodesepB=0.25,linecolor=gray}{00}{2-2}
	\ncline{->,arrowscale=2.0,nodesepA=0.25,nodesepB=0.25,linecolor=gray}{20up}{40up}
	\ncline{->,arrowscale=2.0,nodesepA=0.25,nodesepB=0.25,linecolor=gray}{2-2up}{4-2up}
	\ncline{->,arrowscale=2.0,nodesepA=0.25,nodesepB=0.25,linecolor=gray}{2-2}{40}
	\ncline{->,arrowscale=2.0,nodesepA=0.25,nodesepB=0.25,linecolor=gray}{20}{42}
	\ncline{->,arrowscale=2.0,nodesepA=0.25,nodesepB=0.25,linecolor=gray}{-20}{02}
	\ncline{->,arrowscale=2.0,nodesepA=0.25,nodesepB=0.25,linecolor=gray}{40}{42}
	\ncline{->,arrowscale=2.0,nodesepA=0.25,nodesepB=0.25,linecolor=gray}{4-2}{40}
        \psline[linewidth=1.5pt]{->,arrowscale=1.5}(-4,2.5)(-4,4.0)
        \psline[linewidth=1.5pt]{<-,arrowscale=1.5}(-2,2.5)(-2,4.0)
        \psline[linewidth=1.5pt]{->,arrowscale=1.5}(-0,2.5)(-0,4.0)
        \psline[linewidth=1.5pt]{<-,arrowscale=1.5}(2,2.5)(2,4.0)
        \psline[linewidth=1.5pt]{->,arrowscale=1.5}(4,2.5)(4,4.0)
        \rput[l](4.2,-0.5){along the ray}
        \rput[l](-3.85,3.5){$a_l < 0$}
        \rput[l](-1.85,3.5){$a_l > 0$}
        \rput[l](0.15,3.5){$a_l < 0$}
        \rput[l](2.15,3.5){$a_l > 0$}
        \rput[l](4.15,3.5){$a_l < 0$}
	\end{pspicture}
\caption{Flow of information along a characteristic and wavelength for alternating wavelength couplings.}
\label{fig:stepping}
\end{center}
\end{figure}

To clarify the process of stepping, the flow of information along a characteristic
for alternating wavelength couplings is shown schematically in
Fig.~\ref{fig:stepping}. The arrows mark the direction -- in space and in 
wavelength -- in which the information flows along a characteristic at
wavelength $\lambda$.  It is obvious that the information flows only along the
characteristic and along the wavelength derivative sense determined by the sign of $a_l$.

In the following, the intensities of the $n$th iteration are written with an
additional superscript $(n)$. The iteration step for a specific intensity at
a point $i$ and wavelength $l$ of a characteristic $k$ can then be written for $a^k_{i,l}
\ge 0$ as
\begin{eqnarray}
I^{k,(n+1)}_{i,l} & =  & \left(1 - B^{\mathrm{\diagdown},k}_{i,l} \right)^{-1} \cdot \big( \Delta {I}^{k}_{i,l} +  B^{\mathrm{-},k}_{i,l} I^{k,(n+1)}_{i,l-1} \nonumber \\
 & & \qquad \quad  + \, A^{\mathrm{-},k}_{i,l} I^{k,(n+1)}_{i-1,l-1} + A^{\mathrm{\diagdown},k}_{i,l} I^{k,(n+1)}_{i-1,l} +  C^{\mathrm{-},k}_{i,l} I^{k,(n)}_{i+1,l-1}  \big) \label{eq:explicit_frml_sln} 
\end{eqnarray}
and for $a^k_{i,l} < 0$ as
\begin{eqnarray}
I^{k,(n+1)}_{i,l} & =  & \left(1 - B^{\mathrm{\diagdown},k}_{i,l} \right)^{-1} \cdot \big( \Delta {I}^{k}_{i,l} +   B^{\mathrm{+},k}_{i,l} I^{k,(n+1)}_{i,l+1} \nonumber \\
 & & \qquad \quad   + \, A^{\mathrm{\diagdown},k}_{i,l} I^{k,(n+1)}_{i-1,l}+  A^{\mathrm{+},k}_{i,l} I^{k,(n+1)}_{i-1,l+1} + C^{\mathrm{+},k}_{i,l} I^{k,(n)}_{i+1,l+1} \big), \label{eq:explicit_frml_sln2}
\end{eqnarray}
where all coefficients that vanish have been omitted for the given sign of
$a_l$.

From Eqs.~\ref{eq:explicit_frml_sln}--\ref{eq:explicit_frml_sln2}, it can be
seen that in the case of linear interpolation of the $\hat{S}_l$ source function, the
scheme becomes independent of elements with the iteration order $(n)$ and is
therefore quasi-analytic since it depends only on elements with the same
iteration order. 

In explicit form, the formal solution is given for $a^k_{i,l}
\ge 0$ by:
\begin{eqnarray}
\left(1 - B^{\mathrm{\diagdown},k}_{i,l} \right) I^{k,(n+1)}_{i,l} & =  &  \big( \Delta {I}^{k}_{i,l} +  B^{\mathrm{-},k}_{i,l} I^{k,(n+1)}_{i,l-1} \nonumber \\
 & & \quad  + \, A^{\mathrm{-},k}_{i,l} I^{k,(n+1)}_{i-1,l-1} + A^{\mathrm{\diagdown},k}_{i,l} I^{k,(n+1)}_{i-1,l}  \big) \label{eq:explicit_frml_sln_full} 
\end{eqnarray}
and for $a^k_{i,l} < 0$ by
\begin{eqnarray}
\left(1 - B^{\mathrm{\diagdown},k}_{i,l} \right) I^{k,(n+1)}_{i,l} & =  &  \big( \Delta {I}^{k}_{i,l} +   B^{\mathrm{+},k}_{i,l} I^{k,(n+1)}_{i,l+1} \nonumber \\
 & & \quad   + \, A^{\mathrm{\diagdown},k}_{i,l} I^{k,(n+1)}_{i-1,l}+  A^{\mathrm{+},k}_{i,l} I^{k,(n+1)}_{i-1,l+1}  \big) \label{eq:explicit_frml_sln2_full} 
\end{eqnarray}

The formal solution remains a boundary condition problem,
but it can be solved as an initial value problem at every spatial point. The
solution is then direct and its speed is optimal.

\subsection{Calculation of the formal solution matrix}
\label{sec:03:02}

All non-zero entries in the Matrix $A$ -- also called
characteristic data in the following -- from Eq.~\ref{eq:frml_sln_matrix} must be known
before the formal solution can be calculated.  In the previous algorithm,
the calculation of this characteristic data was parallelised over wavelength to
ensure optimal scaling with the number of processors (see Fig.~$3$ in
\cite{petereddienms3}).  The problem with this approach is that every
process also calculates data it does not need. Furthermore, the data was written to
disk to allow other processes to access the data in case it was needed. This
was developed to allow for small memory demands since the data for a characteristic
could be loaded just before the calculation and deleted afterwards. This strategy
proved to be troublesome for large numbers of wavelength points because the I/O
performance was the limiting factor in the calculations. In non-parallel file
systems, the simultaneous writing and reading of the data files also proved to be
a severe bottle neck. The severity of this problem can be reduced if a
server process does all the I/O and distributes/receives the data to/from the
client processes.

Optimal performance was achieved if the setup for the calculation of
characteristic data was parallelised over characteristics instead. Every process then
calculates the data it will need and stores it directly in memory. This
completely removes the need for I/O and increases the speed by a large factor. Concerns
about load balancing proved to be unfounded.  The calculation of the necessary
data is so fast that the inbalance in the load is not a factor.  This
setup has the disadvantage that for large-scale applications the calculations
cannot be performed on a small number of processors. Because the data files on
disk are absent, the data must be kept in memory at all times and thus a larger
number of processors is needed to perform the domain decomposition effectively. In
practice, this is not a problem since the number of processors is normally chosen to
be large to reduce the overall computing time anyway.

\subsection{Iterative solution of the ALO step}
\label{sec:03:03}

As described in Sect.~\ref{sec:03:00}, the domain decomposition of the ALO
reduces the memory requirements. However, one then must use an iterative
solution for the ALO system. The convergence of iterative solvers  is also 
likely to be very good because the source function in the later iterations will 
already be close to the final solution, whereas direct solutions cannot
take advantage of this. Further iterative solvers are well suited to keeping the 
memory footprint small because they do not have to keep additional data in memory 
besides the linear system.

We implemented GS and Jacobi type solvers to solve 
the linear system of the ALI step: 
\begin{equation}
\label{eq:ali_system}
\left( 1 - \epsilon \, \Lambda^\ast \right) J = J^{\mathrm{FS}} - \epsilon \, \Lambda^\ast  J^{\mathrm{old}}
\end{equation}
where $J$ is the mean intensity and $\epsilon = \frac{\sigma}{\chi}$, where $\sigma$ is the scattering part of the opacity $\chi$.
The mean intensity from the formal solution is $J^{\mathrm{FS}}$ and $J^{\mathrm{old}}$ is the result from the previous ALO step.
The $\Lambda^\ast$-operator is tridiagonal in wavelength, but has the
full spatial bandwidth and its elements are identical to the corresponding $\Lambda$-operator elements.
The three different bands in wavelength are called
$\Lambda^\ast_{-}$,$\Lambda^\ast_{\diagdown}$, and $\Lambda^\ast_{+}$. An example of 
$\Lambda^\ast$ in Eq.~\ref{eq:ali_system} at the wavelength $l$ can then be written as:
\begin{equation}
\Lambda^\ast \left[J_l\right] = \Lambda^\ast_{-} J_{l+1} + \Lambda^\ast_{\diagdown} J_l + \Lambda^\ast_{+} J_{l-1}
\end{equation}

Equation~\ref{eq:ali_system} can then be rearranged into the following form
\begin{eqnarray}
J_{m,l} & = & \left(1 - \epsilon_{m,l} \Lambda^\ast_{\diagdown,m,m,l} \right)^{-1} \cdot \Big( J^{\mathrm{FS}}_{m,l} - \sum_n \epsilon_{n,l} J^{\mathrm{old}}_{n,l}  \Lambda^\ast_{\diagdown,m,n,l} \nonumber \\
 & & \quad - \sum_n \epsilon_{n,l+1} J^{\mathrm{old}}_{n,l+1}  \Lambda^\ast_{-,m,n,l+1}- \sum_n \epsilon_{n,l-1} J^{\mathrm{old}}_{n,l-1}  \Lambda^\ast_{+,m,n,l-1} \nonumber \\
 & & \quad + \sum_{n \ne m} \epsilon_{n,l} J_{n,l}  \Lambda^\ast_{\diagdown,m,n,l} + \sum_{n} \epsilon_{n,l-1} J_{n,l-1}  \Lambda^\ast_{+,m,n,l-1} \nonumber \\
 & & \quad + \sum_{n} \epsilon_{n,l+1} J_{n,l+1}  \Lambda^\ast_{-,m,n,l+1} \Big), \label{eq:ali_element}
\end{eqnarray}
where $n$ and $m$ are indices for the number of layers. Equation~\ref{eq:ali_element} can be readily used in 
the GS and Jacobi iteration schemes.

Because the mean intensity at a wavelength $l$ depends on quantities at the wavelengths $l-1$, $l$, and
$l+1$, the domain decomposition of the ALO in wavelength must be performed blockwise for
the GS method to be applicable. These blocks must overlap with one
wavelength point at the boundaries to minimise communication.

\subsection{Summary of the optimal parallelisation}

\begin{figure}
   \centering
   \includegraphics[width=0.9\hsize]{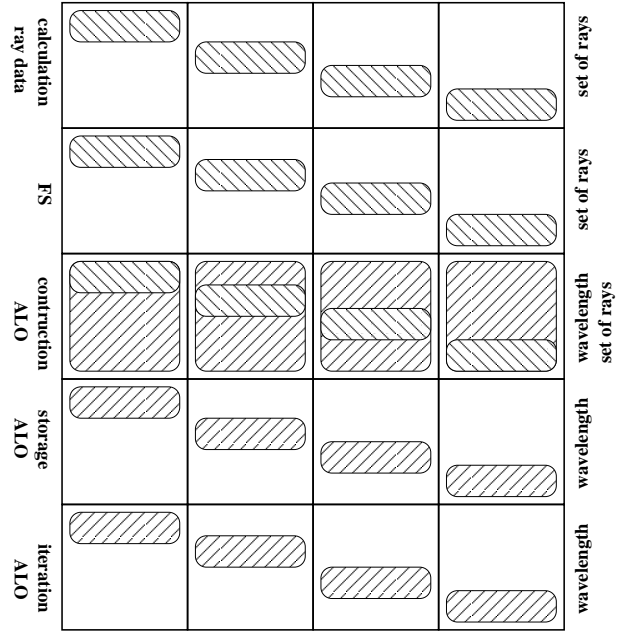}
      \caption{Schematic overview of the parallelisation strategy for 
      four processes. Backward-shaded areas ($\backslash$)
      indicate characteristic-dependent data and forward-shaded areas (/)
      wavelength-dependent data.  }
         \label{fig:002}
   \end{figure}

The parallelisation strategy is the key element for the computation of
large-scale 
applications in the framework of radiative transfer with arbitrary
wavelength couplings. As a summary, the most important aspects are schematically
shown in Fig.~\ref{fig:002}.

\section{Test calculations}
\label{sec:04}

We present the results of test calculations that we performed
to test the new algorithm in terms of speed, scalability, and
by regression testing.

In the following, model calculations have about 20~000 wavelength points and
64~layers unless noted otherwise. This setup would have
been impossible with the old algorithm for radiative transfer as the
memory of an average computing node would have been by far exceeded.
Unless noted otherwise, the computations were performed on 1800 MHz AMD Opteron 244 CPUs
with 4 GB RAM per CPU.

\subsection{Comparison to old algorithm}
\label{sec:04:01}

We compare the new radiative-transfer algorithm with
the methods described in \cite{petereddienms3}.  We show that the improvements
to the formal solution and the ALO step solver described in
Sect.~\ref{sec:03}, are significant.

\subsubsection{Formal solution}
\label{sec:04:01:01}

We compare the speed of two solvers of the formal solution: the
SuperLU solver package and our quasi-analytic solution (see
Sect.~\ref{sec:03:01}). The comparison is  unfair, because the LU decomposition cannot
take advantage of the special character of the matrix. However, the comparison 
demonstrates clearly that the quasi-analytic solution is needed to calculate
large-scale applications in the given framework, since even fast and sophisticated
solvers such as SuperLU are not fast enough to ensure that the calculation is practically feasible.

\begin{figure}
   \centering
   \includegraphics[width=0.9\hsize]{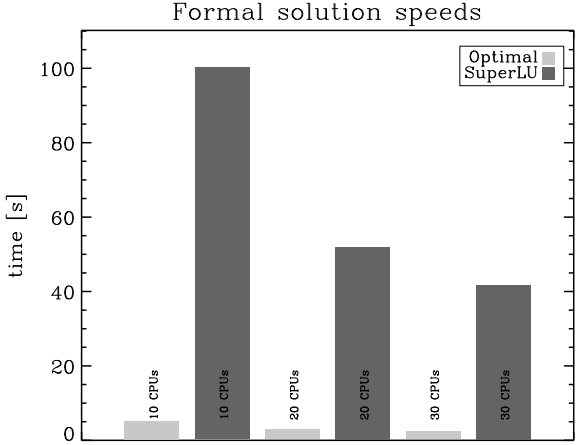}
      \caption{Comparison of the mean times needed for the complete 
      formal solution using the optimal algorithm and SuperLU for
      different numbers of processors.
	      }
         \label{fig:04:01}
   \end{figure}

In Fig.~\ref{fig:04:01}, the mean time needed for a formal solution is shown for
the two solvers and different numbers of processors. 
It is obvious that the optimal solution outperforms the SuperLU package.
The SuperLU solution benefits well from an increase in the number of processors.
This is also true for the optimal solution, but the effect is not as dramatic
because the times are already very short (see Sect.~\ref{sec:04:02:03} for the scalability of
the formal solution).

\begin{figure}
   \centering
   \includegraphics[width=0.9\hsize]{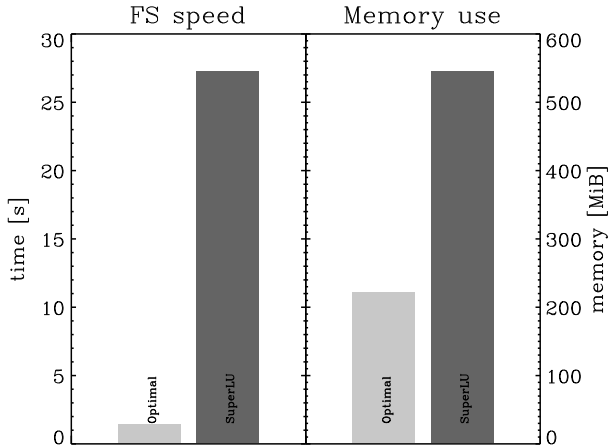}
      \caption{Comparison of the mean time (left) as well as the memory
      footprint (right) needed for the formal solution of the longest
      characteristic using the optimal algorithm and SuperLU.  }
         \label{fig:04:02}
   \end{figure}

In Fig.~\ref{fig:04:02}, the time and memory consumption comparison for the
different solvers is shown for the formal solution along the longest
characteristic in the atmosphere. This allows a more direct comparison of the
solutions. The speed advantage of the
optimal solution is again obvious. The memory footprint of the optimal solution
is also less than half as large as for SuperLU.

The possibility of an analytic solution and the resulting fast formal solution
open up the possibility of calculating radiative transfer in multiple, spatial
dimensions with a characteristic method in the Lagrangian frame, while allowing non-monotonic velocity
fields or other arbitrary wavelength couplings. If the speed of the formal
solution were not optimal, the vast number of 
characteristics would increase the
computation time beyond feasibility.

\subsubsection{Iterative ALO step}
\label{sec:04:01:02}

\begin{figure}
   \centering
   \includegraphics[width=0.9\hsize]{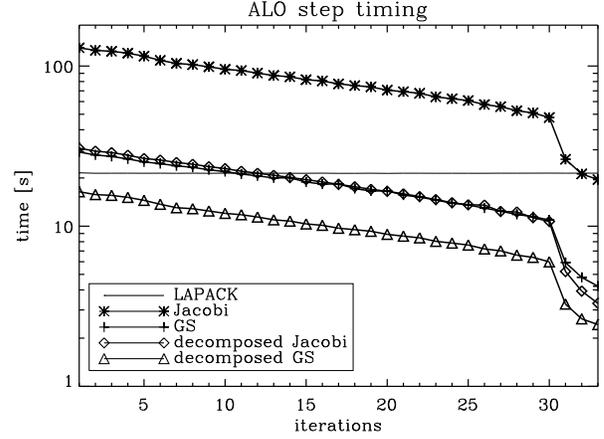}
      \caption{Comparison of the iteration times of the approximate $\Lambda$-iteration for different algorithms.               }
         \label{fig:04:03}
   \end{figure}

In Fig.~\ref{fig:04:03}, we compare the iteration speeds of the ALI for different
algorithms\footnote{The number of wavelength points had to be reduced to
about $10\;000$ in this example due to the increased memory footprint of the
LAPACK solver. Otherwise the calculation would not have been possible on the available
hardware. Furthermore, this test was performed on a
different computer (HLRN-I) than the
other calculations. Hence, direct speed comparisons with other results in this work are
invalid.}. The LAPACK solver has a special role here, since it was the method
of choice in the previous algorithm. As a direct solver the speed is always
the same because it cannot take advantage of the benefits provided by a source function that is already close to
the solution. 

The opposite is true for the iterative solvers. Here we show the results for
Jacobi and GS solvers in serial mode as well as in parallelised versions. The
serial version of the Jacobi solver is significantly slower than the direct
solution. It becomes comparable in efficiency with the LAPACK solver in the
last few iterations because it makes use of the
convergence of the source function.

The serial GS and the parallelised Jacobi solver have similar speeds in the
given example. For the first iterations, they are slower but after about one
third of the iterations they become faster than the LAPACK solver. This results in an
better overall superior performance for the complete ALI.

The parallelised GS solver is even faster than the LAPACK solver from the
beginning and provides the best performance of all solvers, while keeping the
memory requirements to a minimum.

\subsection{Scalability of the new algorithm}
\label{sec:04:02}

By considering scalability, we describe the increased speed and reduced memory footprint of the
algorithm when the strategies from Sect.~\ref{sec:03}  are applied to
different numbers of processors.

In the following, we present the results for the domain decomposition of the ALO,
the ALO construction times, the formal solution speeds, and the ALO iteration
speeds.

\subsubsection{Domain decomposition of the ALO}
\label{sec:04:02:01}

Here we show the drastic effect that the domain decomposition of the ALO has on
the overall memory consumption.

\begin{figure}
   \centering
   \includegraphics[width=0.9\hsize]{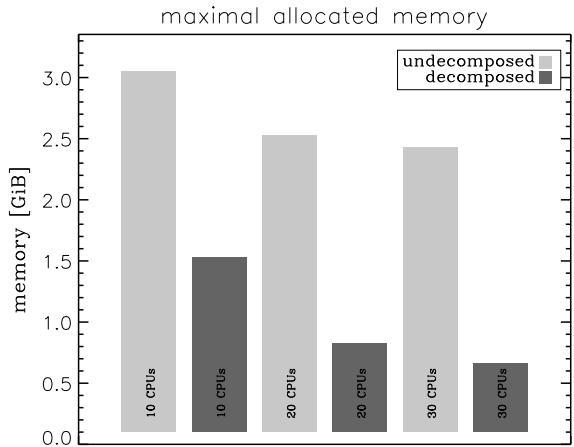}
      \caption{Comparison of the maximum memory used between decomposed and non-decomposed algorithms of the radiative transfer for
      10, 20, and 30 processors.
	      }
         \label{fig:04:04}
   \end{figure}

In Fig.~\ref{fig:04:04}, the maximal allocated memory is shown for
algorithms with and without a decomposed ALO for 10, 20, and 30 processors.
The maximum memory allocated not only includes the ALO but also all other data,
such as the formal solution data and the opacities.

For the non-decomposed as well as decomposed algorithms, the memory footprint
reduces when the number of processors is increased.  In the non-decomposed case,
this is only caused by the reduction in the characteristics data that must be
kept in memory.  In the decomposed case, the smaller amount of ALO data that must
be stored further decreases the memory usage.  It is obvious that the decomposed
algorithm can reduce the memory requirements sufficiently for it to be used
on average current hardware.

\subsubsection{Construction of the ALO}
\label{sec:04:02:02}

\begin{figure}
   \centering
   \includegraphics[width=0.9\hsize]{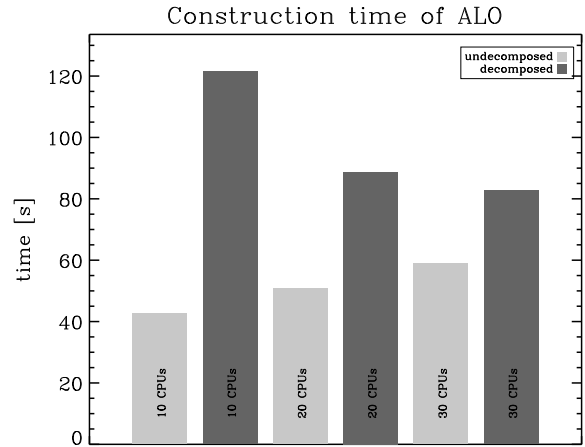}
      \caption{Comparison of the construction times of the approximate
      $\Lambda$-operator between non-decomposed and decomposed setups. }
         \label{fig:04:05}
   \end{figure}

One of the drawbacks of a decomposed ALO is its distributed construction (see
Sect.~\ref{sec:03:00}). Figure~\ref{fig:04:05} shows how the construction
time of the decomposed ALO compares with the construction time of the non-decomposed ALO
for different numbers of processors.

Because the decomposed ALO parts are not calculated by one task alone but
from all formal solution tasks, the construction time increases.  However, the
more formal solution tasks present, the faster the construction becomes. This
can also be seen in Fig.~\ref{fig:04:05}.  For a highly parallelised
calculation with 30 processors, the decomposed construction is only $\approx 1.4$
times slower than in the non-decomposed case. Since the construction of the
ALO must be performed only once during a full ALI, this drawback is not
significant in the light of the smaller memory footprint that the decomposition provides.

\subsubsection{Formal solution}
\label{sec:04:02:03}

\begin{figure}
   \centering
   \includegraphics[width=0.9\hsize]{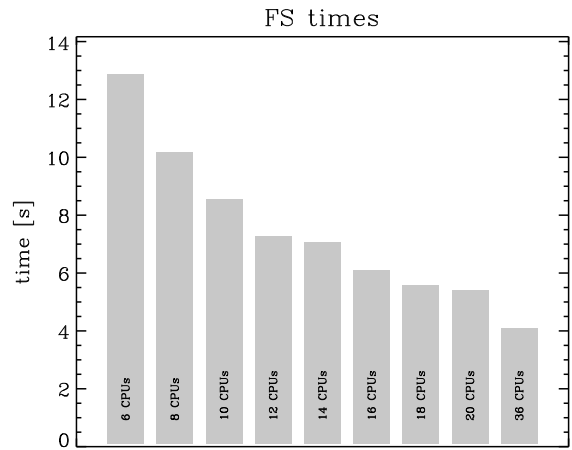}
      \caption{Overview of the timings for the formal solution for different numbers of processors.}
         \label{fig:04:06}
   \end{figure}

The formal solution is the routine that is called the most often during a full ALI solution.
Hence, it is especially important that it be as fast as possible. In
Fig.~\ref{fig:04:06}, the mean times for a full optimal formal solution are shown
for different numbers of processors.

The formal solution speed scales with the number of processors. However, the
scaling is not linear.  The total time needed for a complete formal solution is
limited by the time of the process with the largest characteristics set involved in the computation. 

The overall speed of the local formal
solutions is determined by the number of characteristics and the total number of spatial points along 
characteristics that a task must handle. It follows that an increase in the number of processors produces
a significant increase in speed only if the number of characteristics on the slowest
task is reduced.

The formal solution achieves its optimal speed when there are at least as many
tasks as there are characteristics in the system.

\subsubsection{ALO iteration speeds}
\label{sec:04:02:04}

\begin{figure}
   \centering
   \includegraphics[width=0.9\hsize]{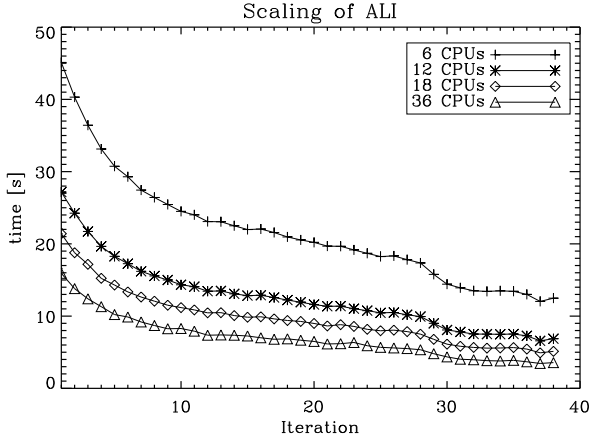}
      \caption{Comparison of the iteration times of the decomposed parallelised GS type approximate $\Lambda$-iteration for a different number of processors.               }
         \label{fig:04:07}
   \end{figure}

After the formal solution, the ALO step is the
next most time-consuming part of the calculation.  In Fig.~\ref{fig:04:07}, the iteration times
for a full ALI are shown for 6, 12, 18, and 36 processors.  In these calculations,
the parallelised GS-type solver has been used.

As the source function comes closer to its final value, the iteration speed
increases for all calculations.  Furthermore, it is clear from Fig.~\ref{fig:04:07}
that the overall speed increases with the number of processors, although the
scaling is clearly non-linear. Nonetheless, the overall computing time of the ALI increases as the number of
processors working on it increases. The asymptotic scaling results from the
need for communication between all ALI tasks after each GS step to
update the iterated solution.

\subsection{Regression test for supernova atmosphere}
\label{sec:04:03}

The formalism and the old algorithm were compared with another
solution for the case of monotonic velocity fields in Paper~I, although the tests
were limited to a toy model. With the new algorithm, we 
are now in the position to calculate full atmosphere models. Hence, it is
instructive to repeat the regression tests in the new framework.
As an example of a large-scale application, we compare the spectra from a
supernova atmosphere calculation with 100 layers and about 20~000 wavelength points. As a reference for the solution of
the radiative transfer in a monotonic velocity-field we used the well tested
and established algorithm described in \cite{peter1992JQSRT}.

The resulting comoving spectra from the old and new algorithm differ for only about
20 wavelength points in the first ten leading digits and the maximal relative differences
are of the order of $10^{-6}$.

This magnitude of difference is at first unexpected, since both radiative-transfer algorithms
use an internal relative convergence criterion of $10^{-8}$. To
understand the remaining differences,
we recall that the iteration procedure differs substantially between the two algorithms.
In the recursive solution, the transfer is solved wavelength by wavelength and the ALI step
will stop as soon as the internal criterion is reached. This is not true for the matrix-based
solution, in which all wavelength points are iterated at the same time. This means that the ALI
will continue to iterate the solution also at those wavelength points, which are already internally converged.

More importantly, the solution from the previous wavelength point is fixed in the recursive scheme and accordingly
the influence of the wavelength coupling does not change in the solution of the radiative transfer at any given 
wavelength point. The opposite
is again true for the matrix-based solution, since the solution is obtained for all wavelength points simultaneously.
This means that the solution converges more successfully at most wavelength points in the matrix-based solution, since the
solution is consistent with the convergence criterion for all wavelengths at the same time.

Taking the points above into account, the overall match of the two solutions is excellent and no residuals were
found in our test calculations.

\section{Conclusion}
\label{sec:05}

We have presented algorithm strategies and details of the solution of the 
radiative-transfer problem in atmospheres with arbitrary wavelength couplings that are
suited to the treatment of large-scale applications. The main aim of the
optimisations of the existing framework has been the reduction of the memory usage
to make the calculations feasible on currently available hardware. 
This has been achieved by domain decomposition of the data and
parallelised code execution. 
In addition, the speed of the formal solution, the calculation of its matrix elements, and the
ALI have all been vastly improved.
The speed of all new algorithms scales with the number of processors used in
the calculations. Although the scaling is non-linear, the overall computation
time is still significantly reduced by an increase in the number of processors.

We are now in a position to calculate large-scale model atmospheres
that include alternating wavelength couplings -- as from non-monotonic velocity
fields or general-relativistic wavelength shifts.

Future possible applications are the velocity profiles of cool stellar winds,
the treatment of partial redistribution, and the calculation of radiative
transfer in shock fronts as in accretion shocks.  However, most promising are the
prospects for the transition of radiative transfer to multiple spatial dimensions. 
Because of the good scaling of the formal solution and the ALO step with the number of
processors, the algorithm strategy can be reused for 3D calculations.

\begin{acknowledgements} This work was supported in part  by SFB 676 from the
DFG, NASA grant NAG5-12127,  NSF grant AST-0707704, and US DOE Grant
DE-FG02-07ER41517. This research used resources of the National Energy
Research Scientific Computing Center (NERSC), which is supported by the Office
of Science of the U.S.  Department of Energy under Contract No.
DE-AC02-05CH11231; and the H\"ochstleistungs Rechenzentrum Nord (HLRN).  We
thank all these institutions for a generous allocation of computer time.
\end{acknowledgements}


\begin{thebibliography}{13}
\expandafter\ifx\csname natexlab\endcsname\relax\def\natexlab#1{#1}\fi

\bibitem[{{Baron} \& {Hauschildt}(2004)}]{petereddienms3}
{Baron}, E. \& {Hauschildt}, P.~H. 2004, \aap, 427, 987

\bibitem[{{Baron} \& {Hauschildt}(2007)}]{peter3d_II}
{Baron}, E. \& {Hauschildt}, P.~H. 2007, \aap, 468, 255

\bibitem[{{Cannon}(1973)}]{cannon1973ApJ}
{Cannon}, C.~J. 1973, \apj, 185, 621

\bibitem[{Demmel {et~al.}(1999)Demmel, Eisenstat, Gilbert, Li, \&
  Liu}]{superlu99}
Demmel, J.~W., Eisenstat, S.~C., Gilbert, J.~R., Li, X.~S., \& Liu, J. W.~H.
  1999, SIAM J. Matrix Analysis and Applications, 20, 720

\bibitem[{Golub \& {Van Loan}(1989)}]{golub89}
Golub, G.~H. \& {Van Loan}, C.~F. 1989, Matrix computations (Baltimore: Johns
  Hopkins University Press)

\bibitem[{{Hauschildt}(1992)}]{peter1992JQSRT}
{Hauschildt}, P.~H. 1992, Journal of Quantitative Spectroscopy and Radiative
  Transfer, 47, 433

\bibitem[{{Hauschildt} \& {Baron}(2004)}]{petereddie2004}
{Hauschildt}, P.~H. \& {Baron}, E. 2004, \aap, 417, 317

\bibitem[{{Hauschildt} \& {Baron}(2006)}]{peter3d_I}
{Hauschildt}, P.~H. \& {Baron}, E. 2006, \aap, 451, 273

\bibitem[{{Hauschildt} \& {Baron}(2008)}]{peter3d_III}
{Hauschildt}, P.~H. \& {Baron}, E. 2008, \aap, 490, 873

\bibitem[{{Knop} {et~al.}(2008){Knop}, {Hauschildt}, \& {Baron}}]{newfrmsol}
{Knop}, S., {Hauschildt}, P.~H., \& {Baron}, E. 2008, inpress in \aap

\bibitem[{{Mihalas}(1980)}]{mihalas80}
{Mihalas}, D. 1980, \apj, 237, 574

\bibitem[{Olson \& Kunasz(1987)}]{frmsol2}
Olson, G. \& Kunasz, P. 1987, Journal of Quantitative Spectroscopy and
  Radiative Transfer, 38, 325

\bibitem[{{Scharmer}(1981)}]{scharmer1981}
{Scharmer}, G.~B. 1981, \apj, 249, 720

\end{thebibliography}
\end{document}